\documentclass[12pt]{iopart}
\usepackage{epsfig}
\usepackage{graphicx}

\begin{document}

\title{The effect of thresholding on temporal avalanche statistics}

\author{Lasse Laurson, Xavier Illa, and Mikko J. Alava}

\address{Department of Applied Physics, Helsinki University of Technology,
FIN-02015 HUT, Finland}

\eads{\mailto{firstname.lastname (at) tkk.fi}}

\begin{abstract}
We discuss intermittent time series consisting of discrete bursts or
avalanches separated by waiting or silent times. The
short time correlations can be understood to follow from
the properties of individual avalanches, while longer time
correlations often present in such signals reflect
correlations between triggerings of different avalanches.
As one possible source of the latter kind of correlations
in experimental time series, we consider the effect of a
finite detection threshold, due to e.g. experimental noise
that needs to be removed. To this end, we study a simple
toy model of an avalanche, a random walk returning to the
origin or a Brownian bridge, in the presence and absence of
superimposed delta-correlated noise. We discuss the properties
after thresholding of artificial timeseries obtained by mixing toy
avalanches and waiting times from a Poisson process. Most of the
resulting scalings for individual avalanches and the composite
timeseries can be understood via random walk theory, except for the
waiting time distributions when strong additional noise is added.
Then, to compare with a more complicated case we study the Manna
sandpile model of self-organized criticality, where some further
complications appear.
\end{abstract}


\maketitle

\section{Introduction}\label{intro}

Many systems in Nature are characterized by an intermittent
avalanche-like response to slow externally applied driving \cite{SET-01}.
Examples range from laboratory scale experiments on magnetic systems
\cite{DUR-05} and e.g. the sound emitted in a fracture test on a piece
of paper \cite{SAL-02}, to various applications in plasma physics and
astrophysics \cite{CHA-01,CHA-03,CHA-99}, as well as in geophysics,
including in particular earthquakes occurring due to the interaction of
slowly moving tectonic plates \cite{GUT-54}. Often the easiest way to
characterize the dynamics of such systems is to record a {\it global}
activity time series $V(t)$, such as the acoustic emission (AE) amplitude
in the paper fracture experiment \cite{SAL-02}, the induced voltage
in a Barkhausen noise measurement \cite{DUR-05} or the AE activity in
martensites \cite{Carrillo1998}. Such signals are
often observed to be composed of apparently distinct bursts or pulses,
which in the limit of slow driving are associated to distinct and
typically spatially localized avalanches of activity occurring in the
system. A typical feature of such bursts or avalanches is that the
statistics of various measures associated to them appear to lack
a characteristic scale, e.g. the avalanche sizes and often also durations are
usually characterized by power law distributions.

Another often made observation is that such avalanche-like
signals appear to exhibit complex temporal correlations.
On shorter time scales, the correlations can be thought to
arise from individual avalanches, and a scaling theory
relating the high frequency power spectra of such signals
to the scaling properties of individual avalanches has been
demonstrated to apply in a number of systems with avalanche
dynamics \cite{KUN-00,LAU-05,LAU-06,ROS-07}.
Longer time correlations are often assumed to be due
to correlations between the triggerings of different avalanches.
Such apparently distinct bursts of activity are often found to
be clustered in time. In the context of earthquakes, events are
usually divided into large main events with smaller fore- and
aftershocks occurring before and after the main shock, respectively.
The rate of aftershocks is typically found to obey Omori's law,
i.e. it decays as a power law in time after the main shock \cite{OMO-95}.
Similar conclusions have been obtained for the foreshocks
\cite{PAP-73,HEL-03}.

Another possibility to detect correlations between different
avalanches is to study the distributions of waiting times
between two consecutive avalanches. If the avalanches are triggered
by an uncorrelated process, one expects the waiting times to obey an
exponential distribution due to the Poisson process -like
random triggering of avalanches \cite{SAN-02}. Consequently, deviations
from this simple form can be interpreted to indicate the
presence of correlations in the triggering process. In a number
of systems, the distributions of waiting times between
avalanches are found to be of power-law type \cite{SAL-02,KOI-07,
COR-04,AMI-06,WHE-02,Guarino2002,Corral_JSTAT}.

Such correlations can have various origins depending
on the physical situation at hand, and it is often difficult
to identify the mechanism operating in a specific experiment.
In systems where the avalanches are triggered by the
external driving (as opposed to situations like thermal creep
in which thermal fluctuations trigger the avalanches), numerical
studies show that the
inter-avalanche correlations can be due to the properties of
the driving signal \cite{SAN-02,BAI-06}. One must also be sure that
a global activity signal is considered: In a {\it local} $V(t)$-signal,
non-exponential waiting times can be observed due to the
spatiotemporal fractal nature of critical avalanches \cite{LAU-04}.
In general, one can define a probability per unit time that
an avalanche is triggered somewhere in the system. If other
mechanisms for the observed correlations such as those mentioned
above can be excluded, a possibility is that the occurrence of an
avalanche somehow intrinsically affects this probability, thus
leading to inter-avalanche correlations.

In this paper we will focus on the effect of a finite detection
threshold necessarily present in any real experimental situation.
The presence of a background, either due to noise or processes
coexisting with the intermittent avalanches in a measured activity time
series $V(t)$, forces one to apply a finite threshold level $V_{th}$,
and define avalanches as the bursts exceeding this threshold
\cite{LAU-06,ROS-07}.
While such ideas have attracted some attention in the literature
\cite{CHR-92,PAC-05,BAI-06_PRL}, neither
theoretically nor experimentally general studies have been carried
out on the effect of thresholding on the avalanche statistics.
Since an avalanche can be defined
as a correlated sequence of activity, breaking the avalanche
into smaller parts by thresholding, the ensuing ``subavalanches''
will be temporally correlated, being part of the same underlying
avalanche. As an attempt to clarify the issues associated to this
effect of thresholding, we consider a simple toy model of the signal
$V(t)$ corresponding to a single avalanche, namely an excursion of
a random walk from the origin. Such a simple model is sufficient to
demonstrate how e.g. the distribution of waiting times arising from
the thresholding process assumes a power law form. We also discuss
the same phenomenology in the stochastic Manna sandpile model of
self-organized criticality \cite{MAN-91}. We also consider the
addition of external noise, and the properties of composite signals
made of toy avalanches and waiting times from a distribution chosen
a priori: how thresholding affects them with or without the noise.
For avalanche sizes and power spectra, there are no real
complications, but for the waiting times we find a mixture of effects
including an apparent power-law regime with a different exponent, when
noise is added.
The paper is organized
as follows: In the next Section, excursions of random walks as a
model of an avalanche signal is considered. The Manna sandpile model
is briefly discussed in Section 3. Section 4 finishes the paper
with conclusions and a summary of future prospects.

\section{The random walk model}

As a conveniently simple toy model illustrating the mechanisms
associated to the thresholding process of avalanche time series,
we consider an excursion of a discrete random walk, obeying
\begin{equation}
\partial_t x = \eta,
\end{equation}
where $\eta$ is white noise with a bimodal distribution
$P(\eta)=1/2(\delta_{\eta,1}+\delta_{\eta,-1})$ and time is updated
in steps of magnitude $\Delta t = 1$. The excursion starts
at $t=0$ from $x=0$, and returns to the origin for the first time
at $t=T$. Such a random walk bridge is then taken to model the $V(t)$-signal
corresponding to a single avalanche of duration $T$.

The statistical properties of such excursions are well known
\cite{RED-01,COL-04,FEL-78}. The first return times, or avalanche 
durations $T$ obey a power law distribution
\begin{equation}
\label{eq:durdist}
P(T) \sim T^{-\tau_T}
\end{equation}
for $T \gg 1$, with $\tau_T=3/2$. Similarly, the distribution of
avalanche sizes $s=\int_0^T V(t) dt$ is given by 
\begin{equation}
P(s) \sim s^{-\tau_s},
\end{equation}
with $\tau_s = 4/3$.
The average shape of the excursion is given
by a semicircle,
\begin{equation}
\label{eq:shape}
V(t,T) = T^{\gamma_{st}-1} f_{shape}(t/T),
\end{equation}
where $f_{shape}(x)=\sqrt{8/\pi}\sqrt{x(1-x)}$ and
$\gamma_{st}=3/2$. The exponent $\gamma_{st}$ relates
the average avalanche size, $\langle s(T)\rangle =
\langle \int_0^T V(t) dt \rangle$ to the duration $T$ as
$\langle s(T)\rangle \sim T^{\gamma_{st}}$.

\subsection{Random walk bridges of a given duration}

\begin{figure}[t!]
\begin{center}
\includegraphics[angle = -90,width = 10cm,clip]{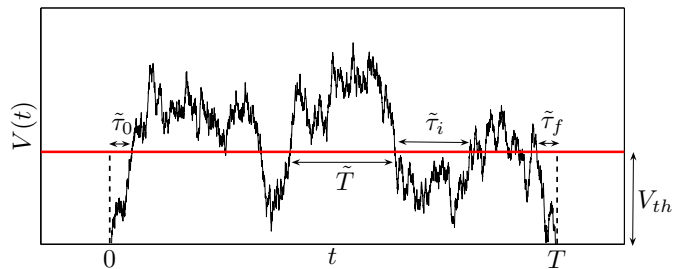}
\end{center}
\caption{An example of subavalanches obtained when a random walk
bridge of given duration $T$ is thresholded with a threshold level $V_{th}$,
with the definitions of the different waiting times (or quiet times)
$\tilde{\tau}_0$, $\tilde{\tau}_i$ and $\tilde{\tau}_f$.}
\label{fig:wt_def}
\end{figure}

First we will focus on the statistical properties of the
avalanches which emerge when random walk bridges of fixed duration $T$
are thresholded as is showed if Fig.~\ref{fig:wt_def}.

\begin{figure}[t!]
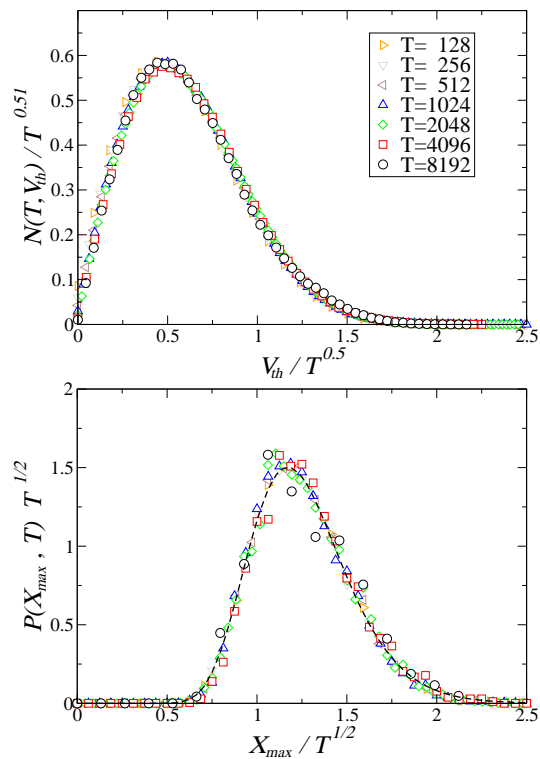

\begin{center}
\includegraphics[width = 7.0cm,clip]{Nas.eps}\\
\includegraphics[width = 7.0cm,clip]{Pxmax.eps}
\end{center}
\caption{Top: The rescaled number of waiting times as a
function of the threshold level $V_{th}$, for different avalanche
durations $T$. Bottom: The rescaled distributions of the
maximum values of the avalanche signal as a function of threshold
level $V_{th}$, for different avalanche durations $T$.
The dashed line is a fit to a log-normal distribution
$f(x;\mu,\sigma)= [1 / (x \sigma \sqrt{2\pi})]
\exp( -(\log(x)-\mu)^2/(2 \sigma^2))$,
with $\mu=0.21$ and $\sigma=0.22$.}
\label{fig:rw_collapse}
\end{figure}

The first quantity of interest is the number of avalanches
$N(T,V_{th})$ as a function of the random walk bridge duration $T$
and the threshold level $V_{th}$. The upper panel of
Fig.~\ref{fig:rw_collapse} shows the number avalanches observed as a
function of the threshold level $V_{th}$, for various durations $T$
of the unthresholded excursions. These can be collapsed onto a
single curve by using the ansatz
\begin{equation}
\label{eq:number}
N(T,V_{th}) = T^{\alpha} f_N (V_{th}/T^{1/2}).
\end{equation}
The value of the exponent $\alpha=0.51 \pm 0.01$ is observed
to be close to $1/2$, a result that can be understood to follow
from the known scaling of the number of zero crossings $N_{zc}(t)$
of a random walk as a function of time, $N_{zc}(t) \sim t^{1/2}$
\cite{FEL-78}.

The lower panel of Fig.~\ref{fig:rw_collapse} displays the distributions
of the maxima of the random walk excursions. Again, a good data
collapse is obtained by using the scaling
\begin{equation}
P(x_{max},T) = T^{-1/2} f_{x_{max}}(x_{max}/T^{1/2}),
\end{equation}
with an apparently universal scaling function $f_{x_{max}}(y)$
that can be fitted well with a log-normal distribution.
The two first moments of this fitted distribution are in agreement with the ones computed in Ref.~\cite{Majumdar2008} using the exact expression.

\begin{figure}[t!]
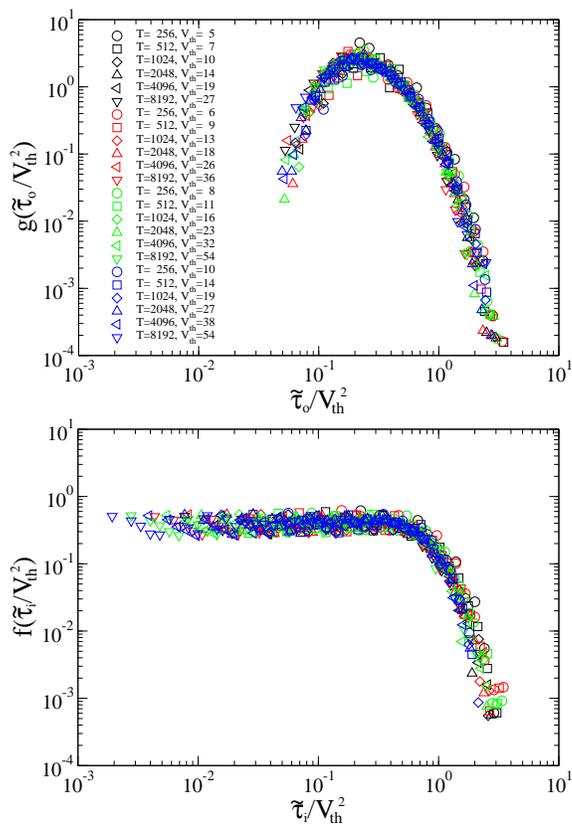

\begin{center}
\includegraphics[width = 7.5cm,clip]{scaling_fT_Ptau2.eps}\\
\includegraphics[width = 7.5cm,clip]{scaling_fT_Ptau3.eps}
\end{center}
\caption{Scaling functions for the initial waiting time distribution
(top) and for the intra-avalanche waiting time distribution (bottom).
Different colours (online version) account for different values of 
$r=V_{th}/T^{1/2} = 0.3,0.4,0.5,0.6$. For larger $r$-values the 
intra-avalanche waiting times $\tilde{\tau}_i$ do not obey Eq. 
(\ref{eq:T_ind}) as for very large $r$ the cut-off due to the finite
avalanche duration becomes important, see Eq. (\ref{eq:T_dep}).}
\label{fig:Ptaus_scaling_fT}
\end{figure}

When a non-zero threshold $V_{th}$ is applied,
one can define three different kinds of waiting times: the initial
waiting time  $\tilde{\tau}_0$, the intra-avalanche waiting times
$\tilde{\tau}_i$,
and the final waiting time $\tilde{\tau}_f$, see Fig.~\ref{fig:wt_def}.
Notice that in this paper we denote all quantities $x$ defined with a
non-zero threshold level $V_{th}$ like $\tilde{x}$, while $x$ stands
for the same thing with $V_{th}=0$. Since for models with a symmetrical 
avalanche shape
$P(\tilde{\tau}_f;T,V_{th})=P(\tilde{\tau}_0;T,V_{th})$, we
focus on the distributions of $\tilde{\tau}_0$ and $\tilde{\tau}_i$.
The former one can be collapsed using the scaling ansatz
\begin{equation}
P(\tilde{\tau}_0;T,V_{th}) \sim V^{-2}_{th} g(\tilde{\tau}_0/V^{2}_{th} ).
\end{equation}
Due to the Markovian and translation invariance properties of the
random walk, $P(\tilde{\tau}_i;T,V_{th})$ is expected to be of a power law
form with the same exponent as $P(T)$ (as $\tilde{\tau}_i$'s are just the
durations of the excursions {\it below} the threshold $V_{th}$),
i.e.
\begin{equation}
\label{eq:T_dep}
P(\tilde{\tau}_i;T,V_{th}) \sim \tilde{\tau}_i^{-3/2}
f(\tilde{\tau}_i/V^{2}_{th}) f_T(\tilde{\tau}_i/T).
\end{equation}
The cut-off scaling follows from the restriction that such
excursions below the threshold are bounded between $0$ and $V_{th}$,
so that a maximum waiting time scaling as $V_{th}^2$ ensues, see Eq.
(\ref{eq:shape}). For random walk bridges of a fixed duration $T$, another
constraint is that $\tilde{\tau}_i < T$. However, in practice the
cut-off scale due to the finite threshold value is reached
first, and the scaling function $f_T(x)$ can be regarded as
a constant. Thus, one can write a $T$-independent form for
the distribution of internal waiting times,
\begin{equation}
\label{eq:T_ind}
P(\tilde{\tau}_i;T,V_{th}) \sim P(\tilde{\tau}_i;V_{th}) \sim
\tilde{\tau}_i^{-3/2} f(\tilde{\tau}_i/V^{2}_{th}).
\end{equation}
Fig.~ \ref{fig:Ptaus_scaling_fT} displays data collapses of
distributions of $\tilde{\tau}_0$ and $\tilde{\tau}_i$. 
The intra-avalanche waiting times $\tilde{\tau}_i$ obey Eq. 
(\ref{eq:T_ind}) only for $r<0.7$ as for very large $r$ the 
cut-off due to the finite avalanche duration becomes important, 
see Eq. (\ref{eq:T_dep}).

\begin{figure}[t!]
\begin{center}
\includegraphics[width = 8cm,clip]{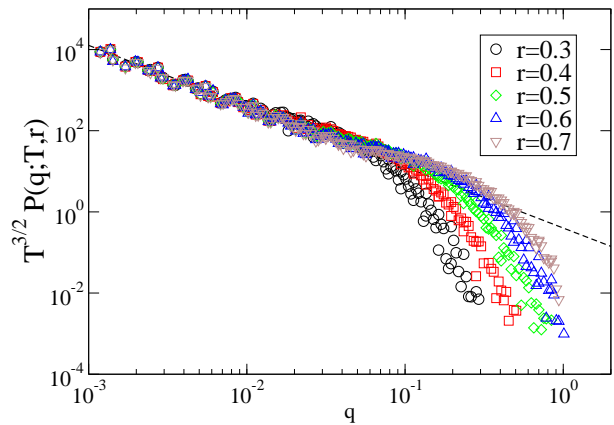}\\
\end{center}
\caption{Scaling functions for the total waiting time distribution
for different values of $r$.}\label{fig:Ptaustot_scaling_fT}
\end{figure}

Next we consider both the initial and the internal waiting times together,
by defining $\tilde{\tau}={\tilde{\tau}_0 \cup  \tilde{\tau}_i}$. The total
distribution of waiting times $\tilde{\tau}$ is then given by
\begin{equation}
 P(\tilde{\tau}; T, V_{th}) = \frac{P(\tilde{\tau}_0;T,V_{th})+(N(T,V_{th})-1)P(\tilde{\tau}_i;T,V_{th})}{N(T,V_{th})}.
\end{equation}
For $T \gg 1$, the number of subavalanches $N(T,V_{th}) \gg 1$, see Eq. 
(\ref{eq:number}) and Fig. \ref{fig:rw_collapse}. In this limit
$N(T,V_{th})-1 \simeq N(T,V_{th})$ and one obtains the total
distribution of waiting times for a fixed duration $T$ and threshold level
$V_{th}$:
\begin{equation}
P(\tilde{\tau}; T, V_{th}) = V^{-2}_{th} \frac{g(\tilde{\tau}/V^{2}_{th})}{f_N (V_{th}/T^{1/2})} T^{-\alpha} +
\tilde{\tau}^{-3/2} f(\tilde{\tau}/V^{2}_{th}).
\end{equation}
Now, using the more natural variables $r=\frac{V_th}{T^{1/2}}$  and
$q=\frac{\tilde{\tau}}{T}$, one can rewrite the above equation as
\begin{equation}
P(q; T,r) = T^{-3/2}
    \left [ \frac{g\left({q}/{r^2}\right)}{r^2 f_N(r)}
  + q^{-3/2} f\left(q/{r^2}\right)   \right ],
\end{equation}
which allows us to collapse the total waiting time distributions for a given
value of $r$ for all durations $T$, as can be seen in Fig.~\ref{fig:Ptaustot_scaling_fT}.
Moreover, the use of these variables also allows us to write the mean value of the
total waiting time distribution as
\begin{equation}
 \left < \tilde{\tau} \right > = T^{1/2} \left [I_1 \frac{r^2}{f_N(r)} + I_2 r\right ],
\end{equation}
where $I_1=\int d\xi \;\;\xi g(\xi)$ and $I_2=\int d\xi \;\; \xi^{-1/2} f(\xi) $.

\subsection{Entire signal}

\begin{figure}[t!]
\begin{center}
\includegraphics[angle=-90,width = 10.5cm]{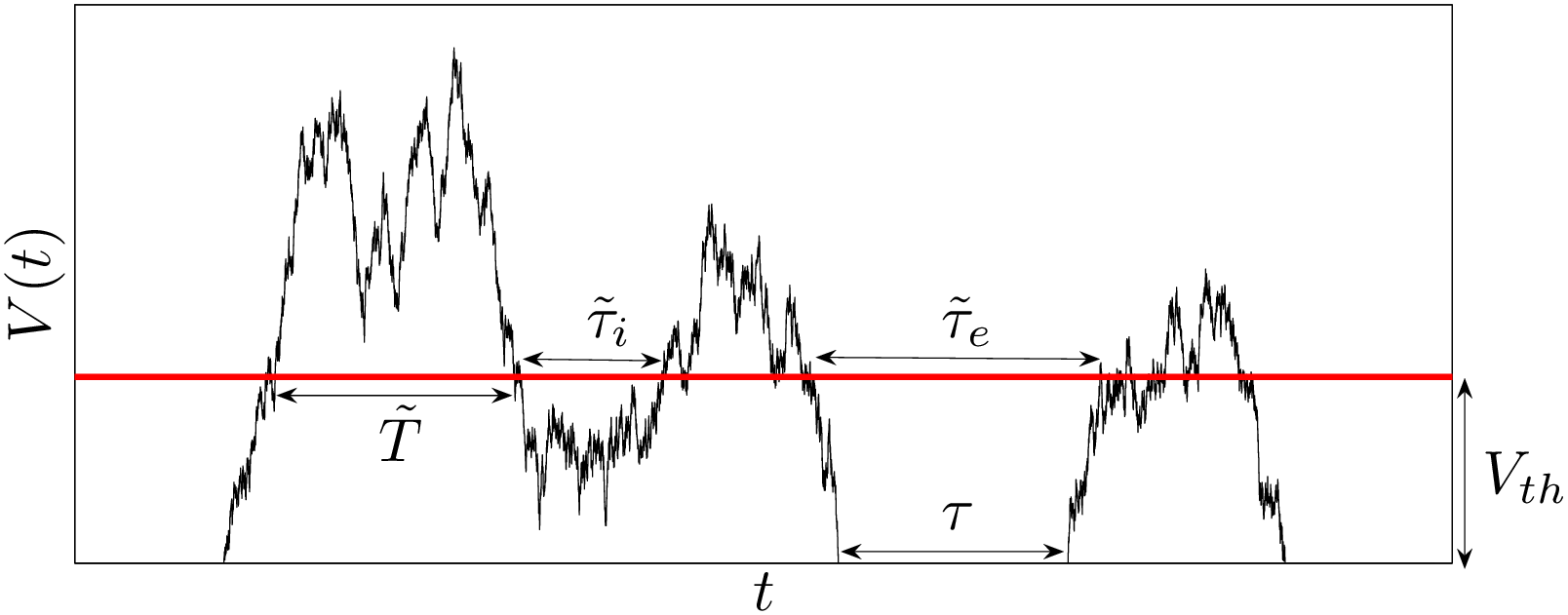}\\
\includegraphics[angle=-90,width = 10.5cm]{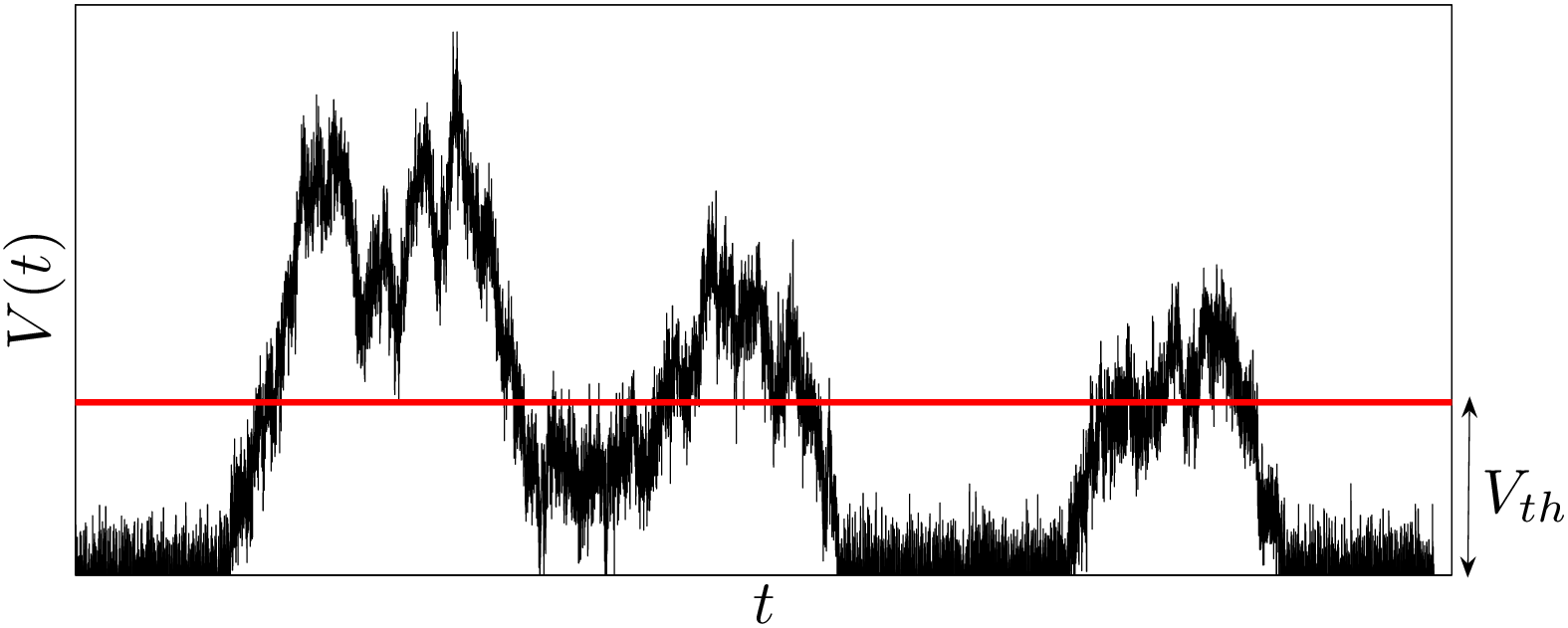}
\end{center}
\caption{Top: An example of a signal $V(t)$ consisting of two
random walk excursions from the origin, showing the definitions
of the avalanche duration $\tilde{T}$ and the waiting
times (or quiet times) $\tilde{\tau}$ after thresholding the signal
with a threshold level $V_{th}$
indicated in the figure by the horizontal line.
Bottom: The same signal as above, but with superimposed Gaussian white
noise. The presence of noise forces one to apply a finite threshold,
but affects also the thresholded avalanches.}
\label{fig:rw_excursion}
\end{figure}

In order to generate an artificial signal,
one can insert exponentially distributed waiting times (corresponding
to uncorrelated triggerings of the avalanches) between
the original (non-thresholded) avalanches and study the evolution of
the waiting time distribution as the threshold level $V_{th}$ is
increased from zero as can be seen in Fig.~\ref{fig:rw_excursion} (top). Also the effect of adding white noise to the random walk signal is considered as can be seen in Fig.~\ref{fig:rw_excursion} (bottom).

\begin{figure}[t!]
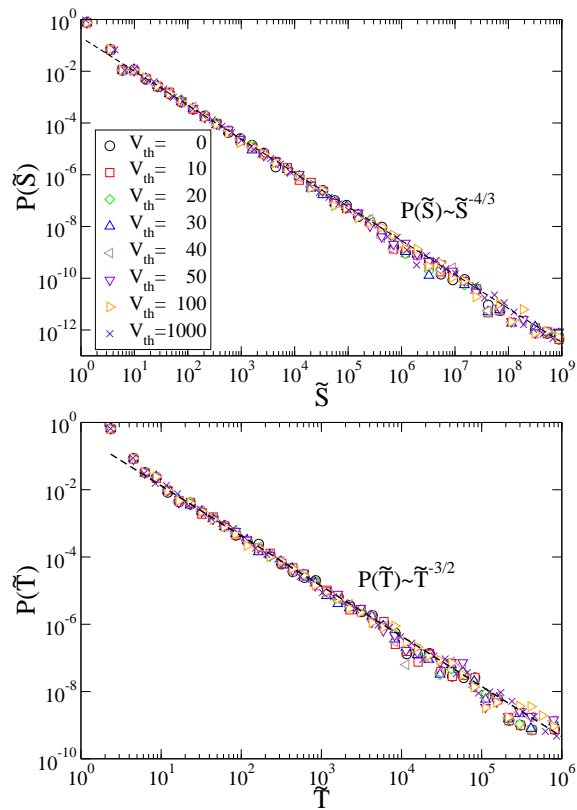

\begin{center}
\includegraphics[width = 7.5cm,clip]{PS_Vth.eps}\\
\includegraphics[width = 7.5cm,clip]{PT_Vth.eps}
\end{center}
\caption{The probability distributions of avalanche sizes $\tilde{s}$
(top) and durations $\tilde{T}$ (bottom) for different threshold levels
$V_{th}$.}
\label{fig:rw_sizedur}
\end{figure}

The total statistics arising from the intra-avalanche properties of
an ensemble of excursions with duration distribution given by Eq.
(\ref{eq:durdist}) at a fixed threshold is obtained as a convolution
over $P(T)$. For instance, one can write for the distribution of
internal waiting times
\begin{equation}
P(\tilde{\tau}_i;V_{th}) = \int P(T) P(\tilde{\tau}_i;T,V_{th}) dT
\sim \tilde{\tau}_i^{-3/2} f(\tilde{\tau}_i/V_{th}^2),
\end{equation}
where the last step follows from the fact that $P(\tilde{\tau}_i;V_{th},T)$
is independent of $T$. Similar considerations apply for the duration
and size distributions of the subavalanches induced by the thresholding
process, such that the thresholded avalanche durations $\tilde{T}$ obey
$P(\tilde{T}) \sim \tilde{T}^{-3/2}$.
Similarly, thresholded avalanche sizes $\tilde{s}=\int_0^T [V(t)-V_{th}] dt$
follow the scaling $P(\tilde{s}) \sim \tilde{s}^{-4/3}$, see
Fig.~\ref{fig:rw_sizedur}.

\begin{figure}[t!]
\begin{center}
\includegraphics[width = 8cm,clip]{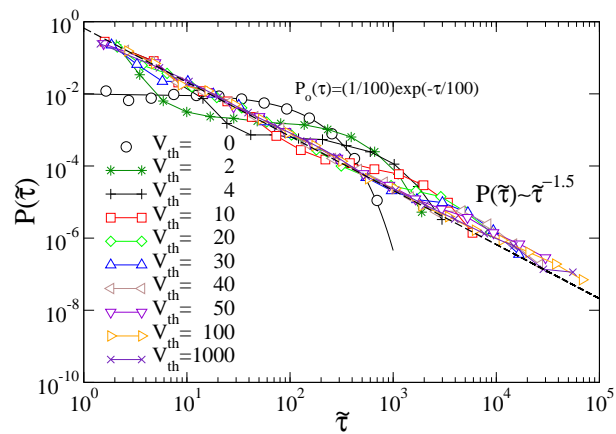}
\end{center}
\caption{The probability distributions of waiting times (or
quiet times) $\tilde{\tau} = \tilde{\tau}_i \cup \tilde{\tau}_e$ 
(see Fig. \ref{fig:rw_excursion}) between avalanches, for different threshold
values, and waiting times between the non-thresholded excursions
drawn from a probability distribution $P_o(\tau)=(1/b)\exp(-\tau/b)$,
with $b=100$.
For $V_{th}=0$ one recovers the exponential $P_o(\tau)$
distribution, while for large enough threshold values the 
distributions become power laws with an exponent $\tau_w \approx 1.5$.}
\label{fig:rw_wait}
\end{figure}


\begin{figure}[t!]
\begin{center}
\includegraphics[width = 7.5cm,clip]{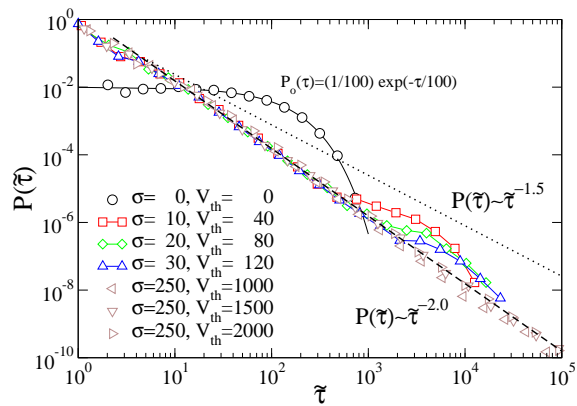}
\end{center}
\caption{The probability distributions of waiting times (or
quiet times) $\tilde{\tau}$ between avalanches with additive
Gaussian white noise, for different threshold values $V_{th}$
and noise strengths $\sigma$. The waiting times between the 
non-thresholded excursions drawn from a probability distribution 
$P_0(\tau)$. The power law exponent
$\tau_w$ changes from the noise free value of $\tau_w = 3/2$ to
$\tau_w \approx 2.0$. Considering the same signal {\it without}
the artificial waiting times (not shown) produces similar distributions,
with the only difference being a decrease in the magnitude of the 
``bump'' in the cut-off due to the artificial waiting times.}
\label{fig:rw_wait_noise}
\end{figure}

When considering the case with exponentially distributed waiting
times $\tau$ inserted between subsequent random walk excursions, and
studying the distributions of all waiting times
$\tilde{\tau} = \tilde{\tau}_i \cup \tilde{\tau}_e$ (see Fig.
\ref{fig:rw_excursion}) as a function of the applied threshold $V_{th}$,
an evolution from an exponential to a power law distribution is
observed as the threshold $V_{th}$ is increased, see Fig.~\ref{fig:rw_wait}.

From an experimental point of view, all $V(t)$-signals are noisy.
In order to treat these real signals, one should add a
minimum threshold to get rid of the background noise but keeping
the maximum amount of information.
Therefore, to study this effect in our artificial signals,
we add a Gaussian white noise $\xi$ with
mean zero and standard deviation $\sigma$ to the $V(t)$-signal and  
consider different thresholds ranging from $V_{th}=4\sigma$ to 
$V_{th}=8\sigma$.
Surprisingly, the exponent appears to change from
$\tau_w = 3/2$ to $\tau_w \approx 2$, see Fig.~\ref{fig:rw_wait_noise}.
This change is because the original (without noise)
waiting times are broken into shorter ones by the noise, 
but we have not managed yet to find a mathematical derivation of
such an exponent.

These results demonstrate how a symmetry between the avalanche
durations and quiet time intervals could give a natural explanation
to observations of power law distributed waiting times in various
experimental situations in which any kind of thresholding process
is applied. To see how such considerations can be generalized to
more realistic systems with non-trivial dynamics, we consider in
the following the same phenomenology in the two-dimensional
stochastic Manna sandpile model of self-organized criticality.

\begin{figure}[t!]
\begin{center}
\includegraphics[angle=-90,width = 10.5cm]{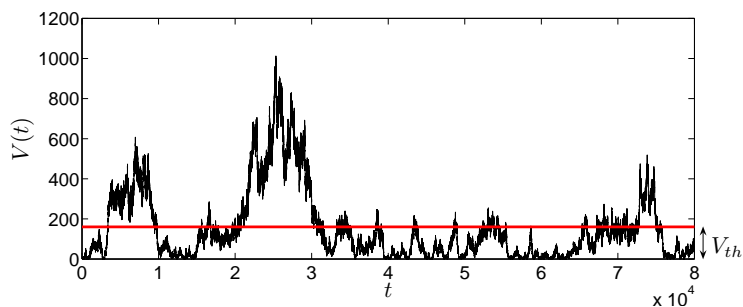}\\
\end{center}
\caption{
An example of a signal $V(t)$ consisting of avalanches
from the $2d$ Manna model with $L=1024$. The horizontal line corresponds
to a threshold level with $V_{th}=160$.
}
\label{fig:manna_signal}
\end{figure}

\section{Manna model}

Out of many possible model systems, we consider here the Manna sandpile
model of self-organized criticality \cite{MAN-91}. This stochastic model
has the
advantage over some other models that it is known to obey simple
scaling, whereas models such as the original sandpile model
introduced by Bak, Tang and Wiesenfeld (BTW) \cite{Bak1987} has been observed to
exhibit multiscaling \cite{TEB-99}. In Ref. \cite{PAC-05}, the effect
of thresholding
on the statistics of quiet time intervals and avalanche durations was
considered. Similarly to our observations above concerning the simple
random walk model, the exponents describing the distributions of the
avalanche durations and quiet time intervals were observed be the same,
but with a value close to $5/3$, different from both the random walk value of
$3/2$ and the value found for the avalanche durations in the BTW model
{\it without} thresholding \cite{LUB-97}. One should notice, however,
that the latter observations regarding the exponent values can be masked
by the multiscaling exhibited by the BTW model \cite{TEB-99}.

The stochastic Manna model is defined on a $d$-dimensional hypercubic
lattice of a linear size $L$, with an integer variable $z_i$ assigned
to each lattice site $i$. A site {\it topples} if the local variable
reaches or exceeds a critical value $z_c=2$. In the toppling process,
two grains will be redistributed from the toppling site to its two
randomly chosen nearest neighbours. During a single time step,
all the sites will be checked and those with $z_i \geq z_c=2$
will be toppled in parallel. Such a toppling can then cause
one of the neighbouring sites of the toppling site to topple during
the next time step. Thus, an avalanche of activity can be triggered
from a single initial toppling event. Such triggerings happen due to
the addition of new grains to random locations in the system at a
slow rate. This external driving is balanced by allowing grains
to leave the system through the open boundaries. This combination of
slow driving and dissipation drives the system to the critical point
of an underlying absorbing phase transition, taking place at a
critical value $\xi_c$ of the grain density $\xi=N/L^d$ \cite{DIC-00,
ALA-02}.

\begin{figure}[t!]
\begin{center}
\includegraphics[width = 7.0cm]{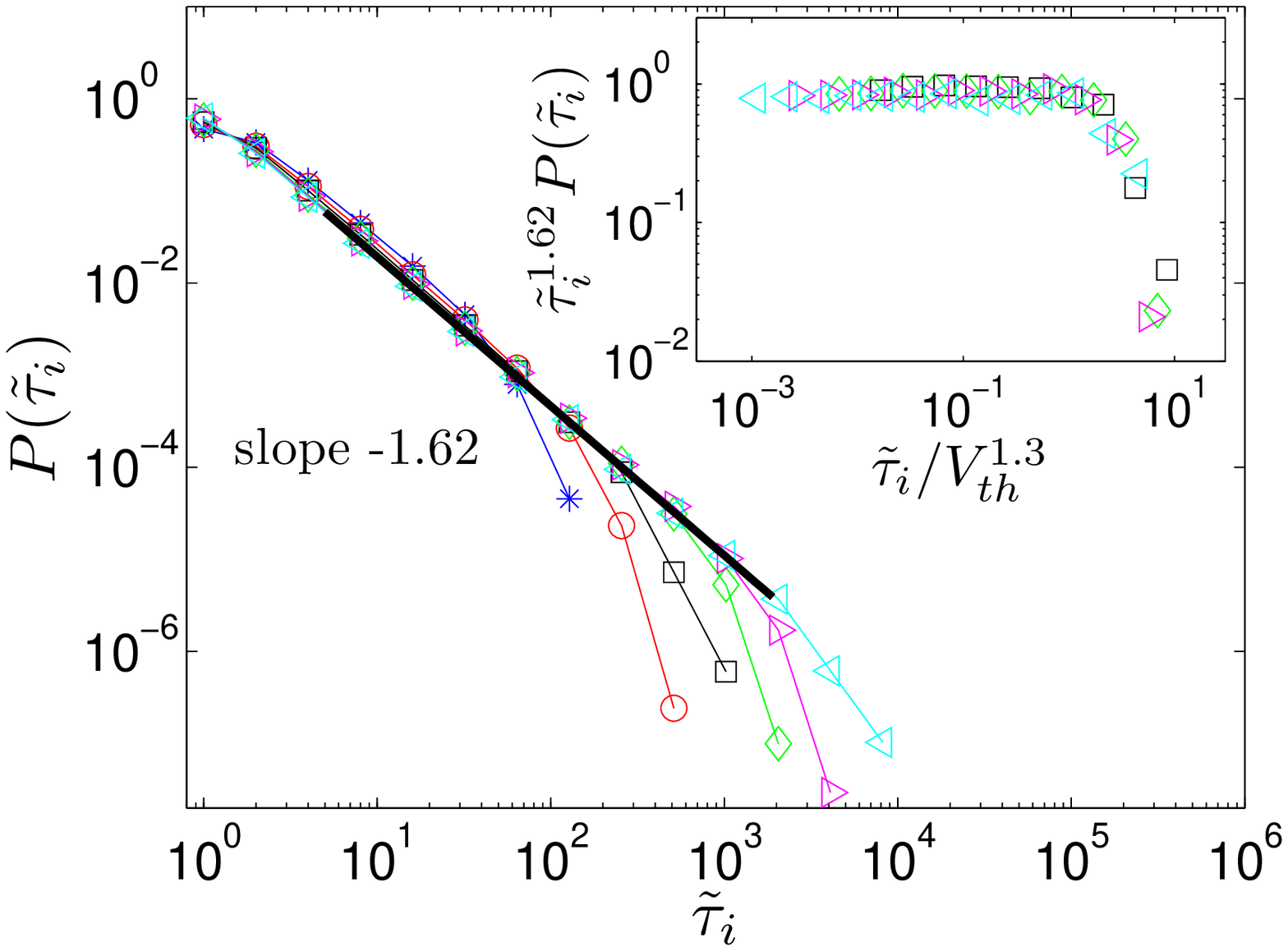} \\
\includegraphics[width = 6.775cm]{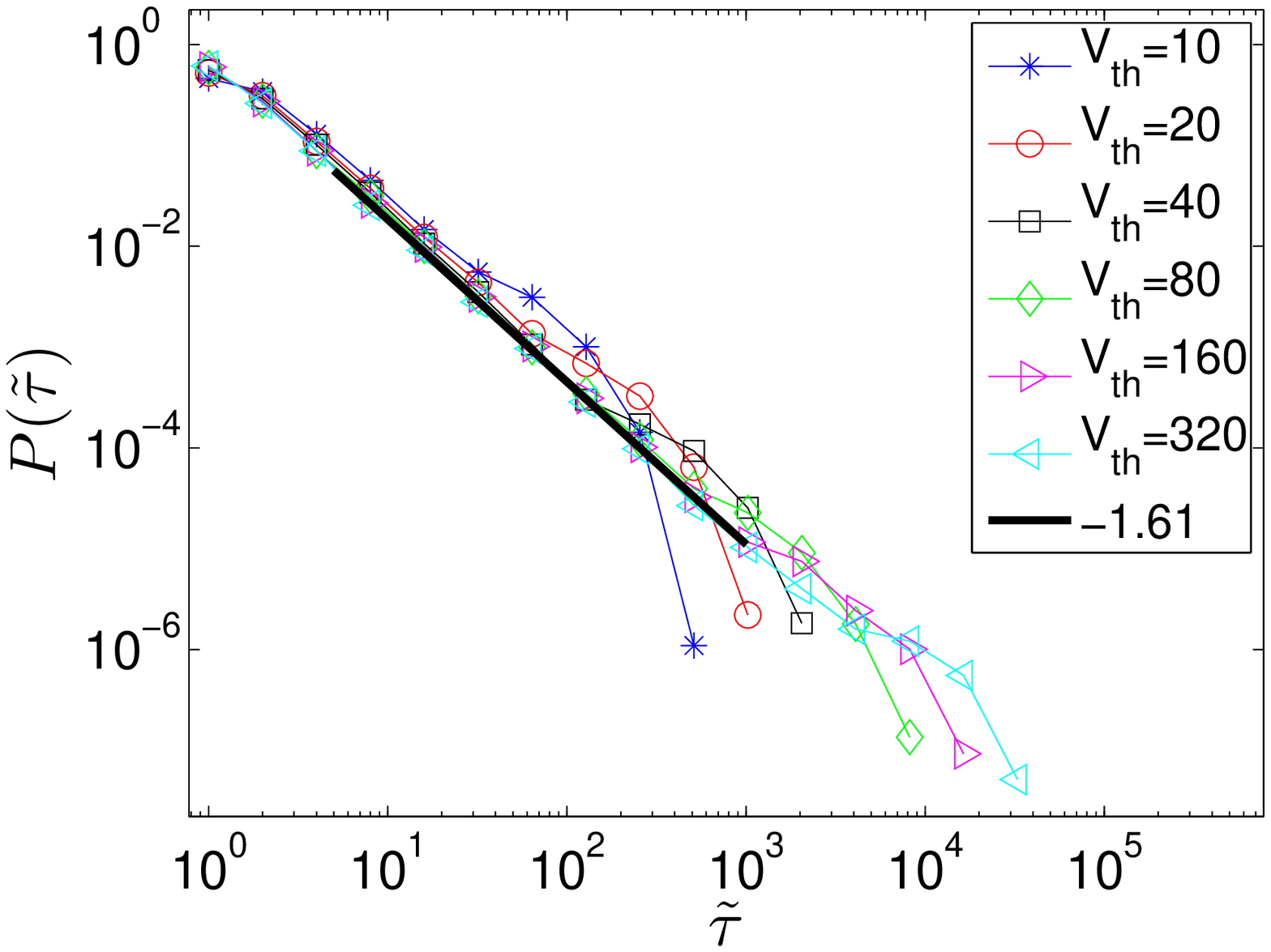}
\end{center}
\caption{Top: The distributions of the intra-avalanche waiting
times $\tilde{\tau}_i$ in the $d=2$ Manna model with $L=1024$ as a function
of the threshold level $V_{th}$. The solid line corresponds to a
power law with an exponent $\tau_w = 1.62$. The inset displays a
data collapse of the distributions for $V_{th} \geq 40$ showing that
the cut-off scales as $\tilde{\tau}_0^* \sim V_{th}^{1.3}$. Bottom: The
distributions of all (intra-avalanche as well as inter-avalanche)
waiting times $\tilde{\tau}$. The solid line corresponds to a power
law with an exponent $\tau_w = 1.61$. }
\label{fig:manna_thr_tws}
\end{figure}

\begin{figure}[t!]
\begin{center}
\includegraphics[width = 7.0cm]{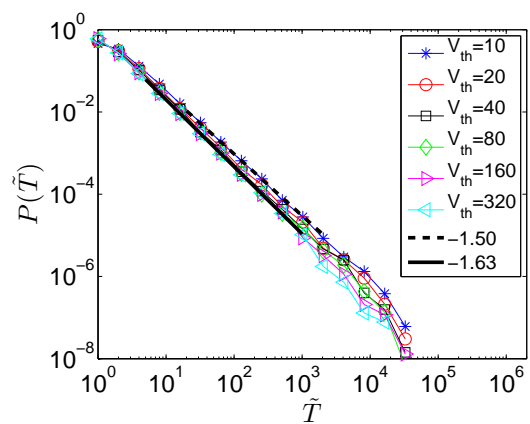}
\end{center}
\caption{The distributions of avalanche durations $\tilde{T}$ in the
$d=2$ Manna model with $L=1024$ as a function of the threshold level
$V_{th}$. The exponent $\tau_T$ appears to evolve from $1.5$ to $1.63$
as the threshold $V_{th}$ is increased.}
\label{fig:manna_durs}
\end{figure}

Here, we consider the two dimensional version of the model, and
study the time series $V(t)$ measuring the number of toppling events as
a function of time, with one parallel update of the lattice defining
the unit of time. Fig. \ref{fig:manna_signal} shows an example of a time
series $V(t)$ from the $2d$ Manna model with $L=1024$. By driving the
system with a slow constant rate, the waiting times between avalanches
without any thresholding will follow exponential statistics. The
non-thresholded avalanche durations are distributed according to a power
law with an exponent $\tau_T \approx 1.5$. The average avalanche size
$\langle s(T) \rangle$ scales with the avalanche duration as
$\langle s(T) \rangle \sim T^{\gamma_{st}}$, with
$\gamma_{st} \approx 1.77$ \cite{LAU-05}. As the threshold level
$V_{th}$ is increased from zero, a power law part starts to emerge to
the internal waiting time distribution, with a cut-off $\tilde{\tau}_i^*$
scaling roughly as
\begin{equation}
\tilde{\tau}_i^* \sim V_{th}^{1/(\gamma_{st}-1)} \approx V_{th}^{1.3},
\end{equation}
The exponent $\tau_{w}$ assumes a value close to the
one observed for the BTW model \cite{PAC-05}, i.e. $\tau_{w}
\approx 1.62 \pm 0.05$, see Fig.~\ref{fig:manna_thr_tws}.
Similar power law scaling is observed by considering the ensemble of
all waiting times, $\tilde{\tau} = \tilde{\tau}_e \cup \tilde{\tau}_i$,
where $\tilde{\tau}_e$ denotes the inter-avalanche waiting times. The
power law exponent remains unchanged from the case where only
$\tilde{\tau}_i$'s are considered, but the inclusion of $\tilde{\tau}_e$'s
has the effect of producing a ``bump''-like cut-off to the distributions.
Interestingly, Fig.~\ref{fig:manna_durs} shows that also the duration
distribution exponent appears to evolve towards a similar value as the
threshold level $V_{th}$ is increased, in agreement with the observations
in the BTW model \cite{PAC-05}.

\begin{figure}[t!]
\begin{center}
\includegraphics[width = 7.0cm]{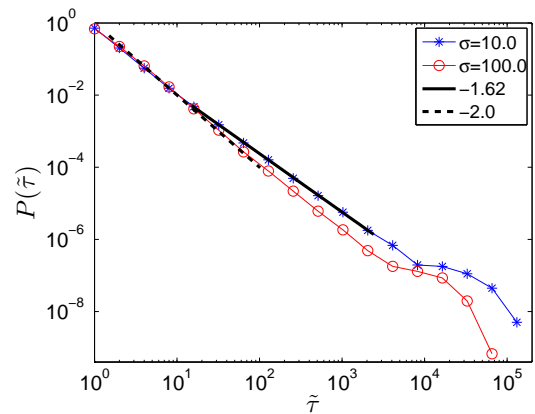}
\end{center}
\caption{The distributions of waiting times $\tilde{\tau}$ in the
$d=2$ Manna model with $L=1024$ and $V_{th}=640$, for two different
strengths $\sigma$ of the additive Gaussian white noise. The exponent
$\tau_w$ appears to change to $\tau_w \approx 2.0$ for short waiting
times, while for waiting times longer than a $\sigma$-dependent
crossover scale, the noise free value $\tau_w \approx 1.62$ persists.}
\label{fig:manna_noise}
\end{figure}

Finally, we consider the effect of noise on the observed waiting time
statistics in the Manna model. By adding Gaussian white noise with
zero mean and standard deviation $\sigma$ to the $V(t)$-signal, we
observe a similar change in the $\tau_w$-exponent as in the case of
the simple random walk model: it assumes a value close to $\tau_w \approx 2$
for waiting times smaller than some crossover scale growing with
$\sigma$. For longer waiting times, the scaling is the same as in the
absence of noise, $\tau_w \approx 1.65$, see Fig.~\ref{fig:manna_noise}.

\section{Conclusions}

The simple random walk model studied here presents a simple
and transparent illustration of the mechanism leading to power law
distributions of quiet times between consecutive subavalanches
emerging from the thresholding process. A key observation here is
that this can be understood to follow from the symmetry between the
random walk excursions above and below the imposed threshold. The
numerical result that such a symmetry appears to be valid also in
the non-trivial case of the Manna sandpile model is an interesting
observation that would deserve further attention.
Also there is an apparent difference in the avalanche duration distribution
exponents in the sandpile models between the unthresholded and
thresholded cases, which is intriguing and not understood. 

The other interesting observation is that additive Gaussian white
noise can have an effect on the observed waiting time statistics.
The apparent change in the power law exponent from the noise-free
value to a value close to -2.0 both in the random walk and Manna
models for high enough noise strength remains to be explained. This
observation may be relevant when noisy experimental signals have to
be thresholded.

Another technique that we have used to study the effect of
thresholding on these artificial signals is the Power Spectrum (PS)
analysis. For the signal composed by random walk excursions the
global behaviour of the PS does not change when thresholding due to
the RW translational invariance, obtaining $P(f)=A/f^2$. The
constant $A$ is decreasing when the threshold is increased due to
the fact that the total energy of the signal decreases. Similar
observations, though less conclusive, can be obtained for the Manna
model: the exponent of the power spectrum does not seem to change
significantly from the non-thresholded case where
$P(f)=A/f^{\alpha}$, with $\alpha = \gamma_{st} \approx 1.77$
\cite{LAU-05}. The exponent $\gamma_{st}$ is related to the
avalanche distribution exponents through
$\gamma_{st}=(\tau_T-1)/(\tau_s-1)$. As we observe an apparent
change in the value of the $\tau_T$-exponent, the $\tau_s$-exponent
should thus experience a corresponding change to keep $\gamma_{st}$
constant. Given the above values for $\tau_T$ and $\gamma_{st}$,
one would thus expect $\tau_s \approx 1.36$, to be compared with 
$\tau \approx 1.28$ for $V_{th}=0$ \cite{LUB-04}. However, due to the limited 
accuracy of the numerical values of these exponents, and to the small 
(if any) change to the PS exponent, we do not pursue this issue.

To discuss the implications of the toy models we have explored we
note that power-law waiting times (and correlations) arise only
after thresholding (single) avalanches. Then the self-affine
character of the single avalanches produces power-law (PL) waiting
times with a particular, model-dependent exponent. For empirical examples
such as earthquakes the question is how to interpret the PL
statistics and correlations in the cascades of aftershocks. Based on
toy models one can envision three reasons for PL waiting times, or
two additional ones beyond such thresholding. First, the system would be
driven externally in a non-Poissonian way such that the PL
statistics ensues. In the case of solar physics this has been argued
to be the case for the magnetosphere, since it might be so that the
solar wind which drives it has such correlations \cite{WHE-00}.
Second, the models we use could be far from the real dynamics, which may
be much more complicated. We represent (in the Manna case, for
instance) the dynamics with a projection to a one-dimensional signal
V(t), but in reality such a trick may produce a V(t) that appears to
have correlations among avalanches while in fact the system evolves
dynamically also when  V(t)=0. A similar mechanism would be that the
system has a memory  which e.g. is influenced by the size of an
avalanche and influences the rate at which subsequent avalanches are
triggered. The usual (e.g. SOC) models do not exhibit such, but this
does not logically exclude the possibility at all.

Finally we point out that in addition to the temporal clustering of
the avalanches, the effect of a finite detection threshold on the
often observed {\it spatial} avalanche clustering could be
considered as well. It is also intriguing as to what would happen to
two-point correlations in spatio-temporal systems, in both time and
space domains.

\ack
Most of this work has been done while visiting Nordita in Stockholm,
and the authors wish to thank the institute for hospitality. LL acknowledges
Satya Majumdar for interesting discussions regarding random walks.
Academy of Finland is thanked for financial support.

\end{document}